\documentclass[twocolumn,superscriptaddress,showpacs,preprintnumbers,amsmath,amssymb,pre]{revtex4}
\usepackage{amsmath}
\usepackage{amssymb}
\usepackage{dcolumn}
\usepackage{graphicx}
\usepackage{bm}
\usepackage{float}

\begin{document}
\title{Protocol Dependence of the Jamming Transition} 
\author{Thibault Bertrand} 
\email{thibault.bertrand@yale.edu}
\affiliation{Department of Mechanical Engineering and Materials Science, Yale University, New Haven, Connecticut, 06520, USA}
\author{Robert P. Behringer}
\affiliation{Department of Physics and Center for Nonlinear and Complex Systems, Duke University, Durham, North Carolina, 27708-0305, USA}
\author{Bulbul Chakraborty}
\affiliation{Martin Fisher School of Physics, Brandeis University, Mail Stop 057,Waltham, Massachusetts, 02454-9110, USA}
\author{Corey S. O'Hern}
\affiliation{Department of Mechanical Engineering and Materials Science, Yale University, New Haven, Connecticut, 06520, USA}
\affiliation{Department of Physics, Yale University, New Haven, Connecticut, 06520, USA}
\affiliation{Department of Applied Physics, Yale University, New Haven, Connecticut, 06520, USA}
\author{Mark D. Shattuck}
\affiliation{Department of Physics and Benjamin Levich Institute, The City College of the City University of New York, New York, 10031, USA}
\affiliation{Department of Mechanical Engineering and Materials Science, Yale University, New Haven, Connecticut, 06520, USA}

\date{\today}

\begin{abstract}
We propose a theoretical framework for predicting the protocol
dependence of the jamming transition for frictionless spherical
particles that interact via repulsive contact forces. We study
isostatic jammed disk packings obtained via two protocols: isotropic
compression and simple shear. We show that for frictionless systems,
all jammed packings can be obtained via either protocol. However, the
probability to obtain a particular jammed packing depends on the
packing-generation protocol. We predict the average shear strain
required to jam initially unjammed isotropically
compressed packings from the density of jammed packings, shape of
their basins of attraction, and path traversed in configuration
space. We compare our predictions to
simulations of shear strain-induced jamming and find quantitative
agreement. We also show that the packing fraction range, over which
shear strain-induced jamming occurs, tends to zero in the large system limit
for frictionless packings with overdamped dynamics.
\end{abstract}

 \maketitle


\section{Introduction}
\label{introduction}
Dry granular materials are composed of macro-sized particles that
interact via repulsive contact forces.  Due to dissipative
interactions between grains, granular materials exist as static
packings in the absence of external driving~\cite{review}.  As a
consequence, granular packings are out-of-equilibrium, and their
structural and mechanical properties depend on the protocol used to
generate them.  Experimental packing-generation protocols include
gravitational deposition~\cite{onoda}, vibration~\cite{vibration},
compression~\cite{behringer}, and shear~\cite{shear_nature,luding}.
Several computational studies have pointed out that the distribution
of jammed packing fractions depends on the compression
rate~\cite{torquato,zhang} and rate at which kinetic energy is removed
from the system~\cite{chaudhuri,schreck}.  In addition, experimental
studies of photoelastic disks have identified key differences between
granular packings generated via isotropic compression and pure
shear~\cite{trush}.

There has been significant work on understanding the
scaling behavior of the elastic moduli and contact number near jamming
onset in model granular packings composed of frictionless
spherical particles generated using isotropic compression~\cite{liu}.
However, there is currently no theoretical understanding of how the
ensemble of static packings and their properties vary with the
protocol used to generate them.  For example, what is  
the difference in the distribution of jammed packings generated 
via isotropic compression versus shear?   

We focus on isostatic jammed packings of frictionless disks generated
via different combinations of isotropic compression and simple shear
and study the distribution of jammed packings as a function of the
path taken through configuration space. A recent study has
distinguished between `compression-only' jammed packings that possess
non-zero pressure and positive shear moduli for some but not all
boundary deformations, and `shear-stabilized' jammed packings
that possess positive shear moduli for all shear
deformations~\cite{dagois-bohy}. We describe the protocol
dependence of {\it compression-only} jammed packings, which are
experimentally realizable~\cite{shattuck} and are relevant for
understanding jamming in systems with frictional interactions~\cite{shen}.

We find several important results.  First, an
exponentially large but finite number of jammed packings with an
isostatic number of contacts $N_c = N_c^{\rm iso}=2N'-1$ (where $N'$
is the number of disks in the force-bearing backbone) exist in
configuration space, defined by the disk positions, packing fraction
$\phi$, and shear strain deformation $\gamma$ of the system
boundaries.  In small systems, nearly all jammed packings can be
enumerated~\cite{shattuck}.  For example, we have shown that isostatic
jammed packings form one-dimensional geometrical families as a
function of shear strain~\cite{gao}.  We will show that the choice of
the packing-generation protocol does not change the ensemble of
isostatic, jammed packings, but instead changes which packings are
visited during particular trajectories through configuration space.
The average properties of jammed packings change for different
protocols because the probabilities for obtaining each jammed packing
varies with protocol.

\begin{figure}
\includegraphics[width=7.8cm]{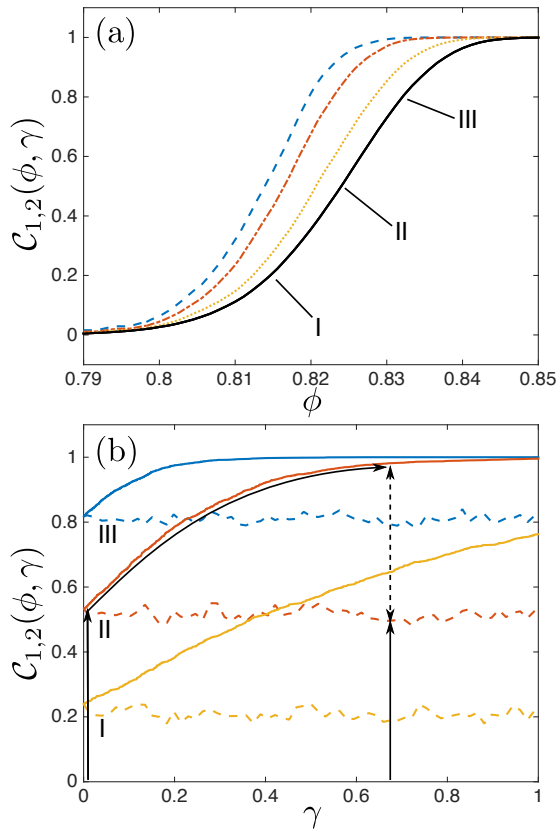}
\caption{Fraction of jammed packings (or cumulative distribution)
${\mathcal C}(\phi,\gamma)$ along different paths in the packing
fraction $\phi$ and shear strain $\gamma$ plane for $N=32$. (a)
${\mathcal C}_{1}(\phi,\gamma)$ for protocol $1$ (isotropic
compression at fixed $\gamma$; solid line) and ${\mathcal
C}_{2}(\phi,\gamma)$ for protocol $2$, {\it i.e.} compression to
$\phi$ followed by shear strain to $\gamma=0.1$ (dotted line),
$0.3$ (dot-dashed line), and $0.5$ (dashed line). I, II, and III
indicate the packing fractions in (b). (b) We show
${\mathcal C}_1(\phi,\gamma)$ (dashed lines) and ${\mathcal
C}_2(\phi,\gamma)$ (solid lines) at fixed $\phi = 0.815$, $0.824$, and
$0.832$ indicated by I-III.  Protocol dependence can be seen in the
difference between ${\mathcal C}_1$ and ${\mathcal C}_2$ evaluated at
the same $\phi$ and $\gamma$, {\it e.g.} at $\phi=0.824$ and
$\gamma=0.67$ as highlighted by the dashed double arrow.  Right
and left solid arrows indicate protocols $1$ and $2$, respectively.}
\label{fig:one}
\end{figure}

\begin{figure}[!h]
\includegraphics[width=7.8cm]{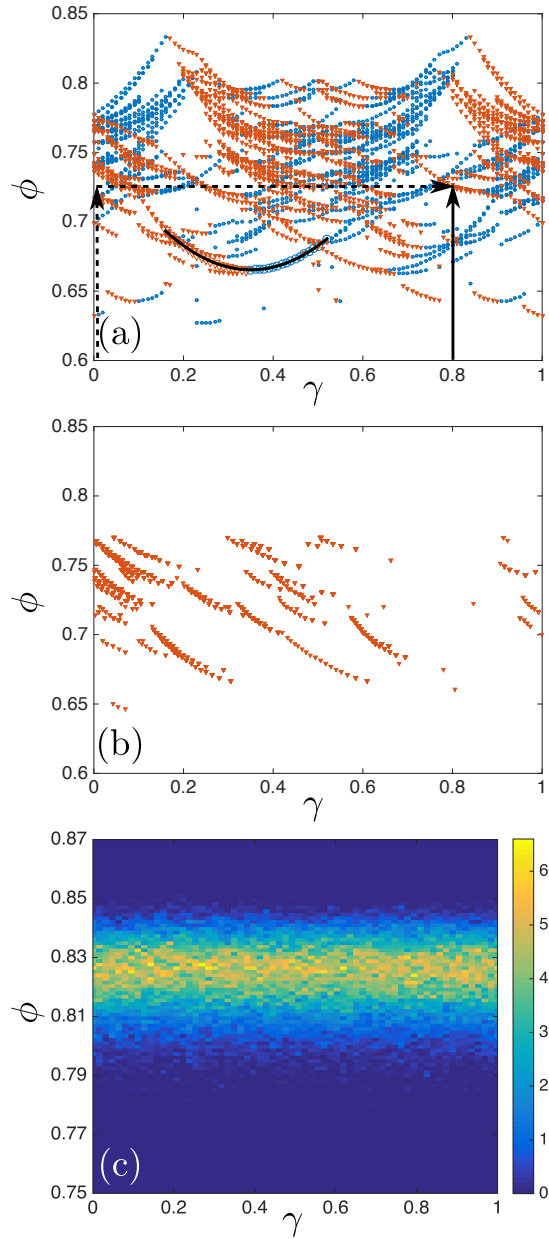}
\caption{(a) Packing fraction $\phi$ versus shear strain $\gamma$ for
all isostatic jammed $N=6$ disk packings. The solid black line obeys $\phi =
A(\gamma-\gamma_0)^2 + \phi_0$ with $A=0.776$, $\phi_0=0.665$, and
$\gamma_0=0.35$. Filled circles (downward triangles) indicate
packings with positive (negative) local slope.  The solid vertical
arrow indicates protocol $1$, and the dashed vertical arrow followed
by the dashed horizontal arrow indicates protocol $2$ that was used to
reach a jammed packing at $\gamma=0.8$ and $\phi=0.725$.
(b) Jammed packing fraction $\phi(\gamma)$ using protocol $2$
at fixed $\phi$ in the range $0.64 < \phi < 0.77$
for $N=6$.  (c) Number of $N=32$ jammed packings at 
each $\phi$ and $\gamma$ (increasing from
dark to light) from protocol $1$.}
\label{fig:two}
\end{figure}

We develop a theoretical description of the protocol dependence of the
distribution of jammed disk packings (Fig.~\ref{fig:one}). We
assume that an initially unjammed system will jam when it encounters
the basin of attraction of a jammed packing as it travels through
configuration space.  The probability to obtain a jammed packing is
determined by two factors: 1) the number density of jammed packings in
configuration space, which is independent of the packing protocol, and
2) the path traveled through configuration space, which depends
on the protocol. Using this framework, we predict the
average shear strain to jam initially unjammed packings at each
$\phi$ and show that the predictions agree with simulations of shear
strain-induced jamming.  Our results indicate that the packing fraction
range, over which shear strain-induced jamming occurs, vanishes in the
large-system limit for overdamped frictionless systems. 

The remainder of the manuscript is organized into three sections. In
Sec.~\ref{methods}, we describe our simulation methods and introduce
the two protocols used to generate isostatic jammed packings. In
Sec.~\ref{results}, we show results concerning the protocol dependence
of the distribution of jammed packings. We then describe a theoretical
model that allows us to calculate the probability to obtain jammed
packings as a function of the path that the system traverses in
configuration space. In Sec.~\ref{conclusion}, we summarize our
results and conclusions.

\section{Methods}
\label{methods}

We study systems containing $N$ frictionless bidisperse disks in a
parallelogram with height $L=1$ in two dimensions (2D) that interact
via purely repulsive linear spring forces with energy scale
$\epsilon$~\cite{epitome}. (Studies of bidisperse frictionless spheres
in three dimensions are included in Appendix~\ref{3D}.)  The mixtures
contain half large and half small particles with mass $m=1$ for both
and diameter ratio $\sigma_L/\sigma_S=1.4$. We employ Lees-Edwards
simple shear-periodic boundary conditions, where the top (bottom)
images of the central cell are shifted to the right (left) by $\pm
\gamma L$ and $\gamma$ is the shear strain~\cite{lees}.  We varied the
system size from $N=6$ to $512$ particles.

Below, we describe results for two protocols to generate jammed
packings in the $\phi$-$\gamma$ plane. (See Fig.~\ref{fig:two} (a).)
Protocol $1$ involves isotropic compression at fixed boundary shape
parametrized by shear strain $\gamma$. We start with random initial
disk positions at $\phi_0<0.5$.  We successively compress the system
by increasing particle radii uniformly in small packing fraction
increments $d \phi$ and minimize the total potential energy per
particle $V/(N \epsilon)$ (at fixed $\gamma$) after each step. Jamming
onset occurs when $V_{\rm max} > V/(N \epsilon) > 0$, with $V_{\rm
max} = 10^{-16}$, or an equivalent threshold on pressure.  For
protocol $2$, we start by isotropically compressing systems (at
$\gamma=0$) to $\phi$, and if the system is unjammed with
$V/(N\epsilon)\ll V_{\rm max}$, we successively apply simple shear
strain to each particle $x_i' = x_i + d \gamma y_i$ in strain steps $d
\gamma < 10^{-3}$ followed by minimization of $V/(N\epsilon)$. We then
identify the total shear strain $\gamma$ at which the system first
jams with $10^{-16} > V/(N \epsilon) > 0$. Protocols $1$ and $2$ 
generate compression-only jammed packings. In Appendix \ref{pure_shear}, 
we also describe results for a third protocol, which is similar to protocol 
$2$, but with simple shear replaced by pure shear. Note
that isostatic jammed packings can also be generated
using stress-controlled packing-generation protocols~\cite{smith}.

\section{Results}
\label{results}

We display the cumulative distributions ${\mathcal
C}_{1,2}(\phi,\gamma)$ of jammed packings from protocols $1$ and $2$ in
Fig.~\ref{fig:one}. In (a), we show that applying shear strain
increases the fraction of jammed packings at each $\phi$, {\it i.e.}
${\mathcal C}_2(\gamma,\phi)$ shifts to lower $\phi$ with increasing
$\gamma$.  In (b), we show ${\mathcal C}_1(\phi,\gamma)$ obtained via
isotropic compression versus boundary shape for
$\phi=0.815$, $0.824$, and $0.832$ (corresponding to ${\mathcal
C}_1(\phi,0) \approx 0.2$, $0.5$, and $0.8$).  $1-{\mathcal
C}_1(\phi,0)$ of the packings are initially unjammed at $\gamma=0$ and
$\phi$, and these jam with increasing
$\gamma$ as shown by the solid lines. By combining different amounts
of shear strain and isotropic compression, the fraction of jammed
packings at a given $\phi$ can be tuned over a wide range, {\it e.g.}
from $0.2$ to $0.8$ for packings at $\phi=0.815$.  These results
emphasize that the distribution of jammed packings depends strongly on
the path through configuration space, {\it e.g.} protocols $1$ and $2$
indicated by the arrows in Fig.~\ref{fig:one} (b).

To understand protocol dependence, we examine the distribution of
jammed packings in the $\phi$-$\gamma$ plane. In Fig.~\ref{fig:two}
(a), we show the packing fraction at jamming onset $\phi$ versus
$\gamma$ for $N=6$ from protocol $1$ (solid vertical arrow in
Fig.~\ref{fig:two} (a)).  We find several striking features. First,
jammed packings occur as geometrical families ({\it i.e.} segments of
parabolas that correspond to jammed packings with the same
interparticle contact network) in the $\phi$-$\gamma$ plane. For
$N=6$, we are able to enumerate nearly all geometrical families over
the full range of $\gamma$~\cite{gao}. When straining an initially
unjammed system toward positive $\gamma$ at fixed $\phi$ ({\it e.g.}
horizontal arrow in Fig.~\ref{fig:two} (a)), it will jam on a
geometrical family with negative slope ($-|d\phi/d\gamma|$). For
negative slopes, continued shear strain leads to overcompression,
whereas for positive slopes, continued shear strain leads to
unjamming. This behavior is shown in Fig.~\ref{fig:two} (b)
for protocol $2$ at fixed $\phi$ in the range $0.64 < \phi < 0.77$ for
$N=6$. Note that any of the jammed packings in Fig.~\ref{fig:two} (b)
from protocol $2$ and defined by $\{{\vec r}_i\}$, $\phi$, and
$\gamma$ can be generated using protocol $1$ with initial condition
$\{{\vec r}_i\}$ and boundary deformation $\gamma$. As a result, we
can generate the same jammed packing at a given $\phi$ and $\gamma$
using different combinations of compression and shear strain. We find
similar behavior to that in Fig.~\ref{fig:two} (a) and (b) for larger
$N$, except that the parabolic segments in $\phi(\gamma)$ become
smaller and more numerous, and thus geometrical families more densely
populate configuration space (Fig.~\ref{fig:two} (c)). The number of
jammed packings (at a given $\phi$ and $\gamma$) becomes independent
of $\gamma$ for $N\ge32$.

\begin{figure}
\includegraphics[width=7.8cm]{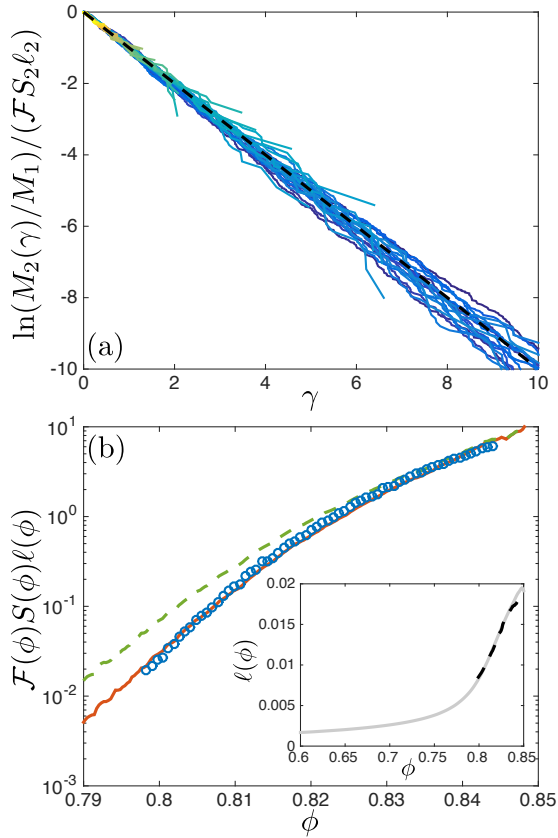}
\caption{(a) Natural logarithm of the fraction of unjammed packings
(normalized by the $\phi$-dependent decay factor, ${\mathcal
F}(\phi)S_2(\phi) \ell_2(\phi)$), during protocol $2$ at fixed $\phi$
in the range $0.798 < \phi < 0.844$ (solid lines) for $N=32$.  The
dashed line has slope $-1$. (b) Comparison of $-d \ln
[M_2(\phi,\gamma)/M_0]/d\gamma$ (open circles) from protocol $2$ and
$-d\ln[M_1(\phi)/M_0]/d\phi$ from protocol $1$ with $S_2(\phi) \propto
S_1(\phi) \ell_1(\phi)$ (solid line) or $S_2(\phi) \propto S_1(\phi)$
(dashed line) for $N=32$. The inset shows the distances
$\ell_1(\phi)$ and $\alpha \ell_2(\phi)$ (with $\alpha \approx 5.5$)
traversed in configuration space during protocols $1$ (solid line) and
$2$ (dashed line) for $N=32$.}
\label{fig:three}
\end{figure}

\begin{figure}
\includegraphics[width=7.8cm]{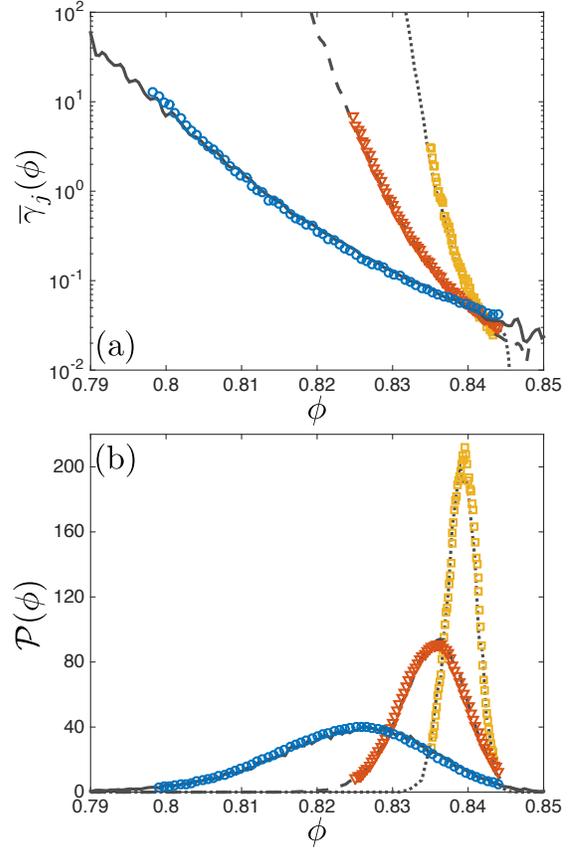}
\caption{(a) Average shear strain ${\overline \gamma}_j(\phi)$
to jam an initially unjammed configuration at $\phi$ and
$\gamma= 0$ (protocol $2$) for $N = 32$ (circles), $128$ (triangles),
and $512$ (squares). We compare ${\overline \gamma}_j(\phi)$ from
protocol $2$ to $1/({\mathcal F}(\phi) S_1(\phi) \ell^2_1(\phi))$
(Eqs.~\ref{average_strain} and~\ref{average_strain2}) from
protocol $1$ for the same system sizes: $N=32$
(solid line), $128$ (dashed line), and $512$ (dotted line). (b)
Distribution of jammed packing fractions $\mathcal{P}_1(\phi)$
from protocol $1$ for $N=32$ (solid line), $128$
(dashed line), and $512$ (dotted line) compared to predictions
obtained from ${\mathcal P}_1(\phi)= -M_0^{-1}dM_1(\phi)/d\phi$ with
$M_1(\phi)$ given by Eq.~\ref{n0} and ${\mathcal F}(\phi)S_1(\phi)
l_1(\phi)$ given by Eq.~\ref{average_strain2} using the measured value
of ${\overline \gamma}_j$ for $N = 32$ (circles), $128$ (triangles),
and $512$ (squares).}
\label{fig:four}
\end{figure}

We develop a theoretical model using an
analogy with absorption problems to calculate the probability to
obtain isostatic jammed packings as a function of the path that the
system traverses in configuration space. In principle, the number density of 
jammed packings ${\mathcal F}$ depends on the $2N$ coordinates of the 
disks, $\phi$, and $\gamma$, but not the packing-generation protocol. 
After integrating over the $2N$ disk coordinates, ${\mathcal F}$
is a function of $\phi$ and $\gamma$. However, we assume that the 
number density ${\mathcal F}(\phi)$ is only a function of $\phi$ since 
${\mathcal F}$ becomes independent of $\gamma$ in the 
large-$N$ limit (Fig.~\ref{fig:three} (c)). We imagine that a
one-dimensional trajectory $\mathcal{L}(\phi,\gamma)$ through
configuration space will encounter the basin of
attraction~\cite{ashwin} of a jammed packing with a probability
${\mathcal F}(\phi)S(\phi) d\mathcal{L}$ during a step of size $d
\mathcal{L}$ in configuration space, where $S(\phi)$ is the
$2N-1$-dimensional cross-section of the basin of attraction of a
jammed packing perpendicular to $d \mathcal{L}$.  

Thus, for protocol $1$ with trajectories only along $\phi$, the
decrease in the number of unjammed packings $d M_1(\phi)$ (or
equivalently, increase in the number of jammed packings) during a
compression step $d \phi$ is
\begin{equation}
\label{diffeq}
d M_1(\phi)=-M_1(\phi)
{\mathcal F}(\phi)S_1(\phi) \ell_1(\phi)d \phi, 
\end{equation}
where $d \mathcal{L}=\ell_1(\phi) d\phi$, $\ell_1(\phi)$ is the
distance in configuration space traversed during step $d \phi$ at
$\phi$, and $S_1(\phi)$ is the average cross section
for protocol $1$.  Eq.~\ref{diffeq} can be solved for the
number of unjammed packings at $\phi$ during protocol $1$:
\begin{equation}
\label{n0}
M_1(\phi)=M_0 \exp\left[-\int^\phi_{\phi_0} {\mathcal F}(\phi')S_1(\phi') \ell_1(\phi')d\phi'\right],
\end{equation}  
where $M_0$ is the number of unjammed packings at $\phi_0$.  For
protocol $2$, with trajectories only along $\gamma$, we obtain a
similar expression for the number of unjammed packings: $d
M_2(\phi,\gamma)/d\gamma=-M_2(\phi,\gamma) {\mathcal F}(\phi)S_2(\phi)
\ell_2(\phi)$, where $M_2(\phi,\gamma)=M_1(\phi) \exp[-{\mathcal
F}(\phi) S_2(\phi) \ell_2(\phi) \gamma]$, $S_2(\phi)$ is the average
cross section for protocol $2$, and $\ell_2(\phi)$ is the distance
traversed in configuration space for each shear strain step $d
\gamma$.

Fig.~\ref{fig:three} (a) shows that the fraction of unjammed packings
$M_2(\phi,\gamma)/M_1(\phi)$ decays exponentially with $\gamma$ during
protocol $2$ at each $\phi$. This result emphasizes that we can calculate the 
product ${\mathcal F}(\phi)S(\phi) \ell(\phi)$ without enumeration of 
all jammed packings by measuring the decrease in the number of 
jammed packings with increasing shear strain. In the zeroth order
approximation, the cross section $S(\phi)$ is independent of
the path in configuration space and the distance $\ell$ traveled
during each $d\phi$ or $d\gamma$ step is constant. In
Fig.~\ref{fig:three} (b), we compare ${\mathcal F}(\phi) S(\phi) \ell$
from protocol $2$ with the similar quantity
$-d\ln[M_1(\phi)/M_0]/d\phi$ from protocol $1$ and find qualitative
agreement.

We then independently measured $\ell_1(\phi)$, defined by the
accumulated distance in configuration space between the initial
packing at $\phi$ and relaxed packing at $\phi+d \phi$, for 
protocol $1$. We performed similar measurements for
$l_2(\phi)$, which gives the accumulated distance in configuration
space between the initial packing at $\gamma$ and relaxed packing
at $\gamma+d\gamma$ for protocol $2$. We show that the two 
are proportional to each other, $\ell_1(\phi) = \alpha \ell_2(\phi)$
with $\alpha \approx 5.5$, in the inset of Fig.~\ref{fig:three} (b).
By calculating $-d \ln[M_1(\phi)/M_0]/d\phi = {\mathcal F}(\phi)
S_1(\phi) \ell_1(\phi)$ for protocol $1$, we
can compare ${\mathcal F}(\phi) S_1(\phi) \ell_1(\phi)$ to the similar
quantity, ${\mathcal F}(\phi) S_2(\phi) \ell_2(\phi)$, for protocol
$2$. In this case, we assume that the cross
section depends on the path in configuration space, {\it e.g}
isotropic compression increases the overlaps of all interparticle
contacts, while shear strain increases some but decreases
others. In Fig.~\ref{fig:three} (b), we show excellent agreement for
${\mathcal F}(\phi) S_{1,2}(\phi) \ell_{1,2}(\phi)$ for protocols $1$
and $2$ for $N=32$ provided we assume that $S_2(\phi) \propto
S_1(\phi) \ell_1(\phi)$, and find similar quantitative agreement
for all system sizes studied. (Additional details of the theoretical model
are included in Appendix~\ref{theory}.) Independent measurements of
$S_{1,2}(\phi)$ will be performed in future studies.

We now use the theoretical description of the protocol-dependent
probability to jam to predict the average shear strain required to jam
an initially unjammed isotropically compressed configuration at $\phi$
and $\gamma=0$:
\begin{eqnarray}
\label{average_strain}
{\overline \gamma}_j  (\phi)& = & \int_0^{\infty} \gamma
\frac{M_2(\phi,\gamma)}{M_1(\phi)} d\gamma = \frac{1}{ {\mathcal
F}(\phi)S_2(\phi)\ell_2(\phi)} \\
\label{average_strain2}
& \simeq & \frac{\alpha}{ {\mathcal F}(\phi) S_1(\phi) \ell^2_1(\phi)}.
\end{eqnarray} 
In Fig.~\ref{fig:four} (a), we show that the prediction for
${\overline \gamma}_j(\phi)$, obtained from measurements of ${\mathcal
  F}(\phi) S_1(\phi) \ell^2_1(\phi)$ using isotropic compression,
agrees with simulations of shear strain-induced jamming. We find that
${\overline \gamma}_j$ grows rapidly with increasing system size and
only packings with $\phi \gtrsim 0.84$ are jammed in the large-system
limit~\cite{ning}.  (We find similar results for applied pure shear in
Appendix~\ref{pure_shear}.)  We also calculate the distribution
${\mathcal P}_1(\phi)$ of jammed packing fractions (for isotropic
compression) using data from protocol $2$. In Fig.~\ref{fig:four} (b),
we show that ${\mathcal P}_1(\phi)$ from protocol $1$ and ${\mathcal
  P}_1(\phi)= -M_0^{-1}dM_1(\phi)/d\phi$ with $M_1(\phi)$ given by
Eq.~\ref{n0} and ${\mathcal F}(\phi)S_1(\phi)\ell_1(\phi)$ given by
Eq.~\ref{average_strain2} (using the measured value of ${\overline
  \gamma}_j$) collapse for all system sizes studied. The width of
${\mathcal P}_1(\phi)$ for isotropic compression narrows as
$1/N^{\lambda}$ with $\lambda \approx 0.55\pm 0.05$ and the peak
approaches $\phi_{\rm rcp} \approx 0.84$ in the large-system
limit~\cite{ning}.
 
 \section{Conclusion}
 \label{conclusion}
In this manuscript, we developed a theoretical description for jamming
onset that allows us to predict the fraction of isostatic jammed
packings that occur at $\phi$ and $\gamma$ in terms of the path
traversed in configuration space.  This framework provides predictions
for the average shear-strain required to jam initially unjammed
packings produced by isotropic compression, which agree quantitatively
with simulations of strain-induced jamming in two- and
three-dimensional systems subjected to simple and pure shear.  In
particular, we showed that the packing fraction range, over which
strain-induced jamming occurs, shrinks to zero in the large-system
limit for frictionless systems with overdamped dynamics. In future
studies, we will investigate the role of static friction in
stabilizing strain-induced jamming of dilute granular
packings~\cite{shear_nature}.

\begin{acknowledgements}
We acknowledge support from the W. M. Keck Foundation Grant No. DT061314
(T.B., R.B., B.C., and C.S.O.), and National Science Foundation
(NSF) Grant Nos. CBET-0968013 (M.D.S.) and DMR-1409093 (B.C.). We also
acknowledge support from the Kavli Institute for Theoretical Physics
(NSF Grant No. PHY-1125915), where some of this work was
performed. This work benefited from the facilities and staff of the
Yale University Faculty of Arts and Sciences High Performance
Computing Center and NSF Grant No. CNS-0821132 that in part
funded acquisition of the computational facilities.
\end{acknowledgements}

\appendix
\section{3D bidisperse packings}
\label{3D}

In this section, we present our studies of compression and
shear-strain induced jamming of 3D bidisperse spheres, which are
qualitatively similar to the results for 2D bidisperse systems
presented in the main text. We presented results in the main text on the distribution of jammed
packing fractions ${\cal P}(\phi)$ and average shear strain ${\overline
\gamma}_j$ required to induce jamming in an originally unjammed
configuration for systems composed of bidisperse disks in two spatial
dimensions ($d=2$).  However, these results apply more generally than
simply to two-dimensional packings of disks.  The theoretical analysis
in the main text described trajectories in the $dN$-dimensional
configuration space in which jammed packings exist, where $d$ is the
spatial dimension.  The 2N-dimensional configuration space is already
large, and thus we expect qualitatively the same results for
three-dimensional (3D) sphere packings, which exist in a configuration
space that is only $50\%$ larger, as we found for two-dimensional
systems.

We studied systems containing $N$ frictionless bidisperse spheres in a parallelepiped with sides of
length $L = 1$ that interact via purely repulsive linear spring
forces. The bidisperse mixtures contain half large and half small
particles, both with mass $m=1$, and diameter ratio $\sigma_L/\sigma_S
= 1.4$.  As in 2D, we employ Lees-Edwards simple shear-periodic
boundary conditions, where the top (bottom) images of the central cell
are shifted to the right (left) by $\gamma L$, where $\gamma$ is the 
simple shear strain.

\begin{figure}[!h]
\includegraphics[width=8.5cm]{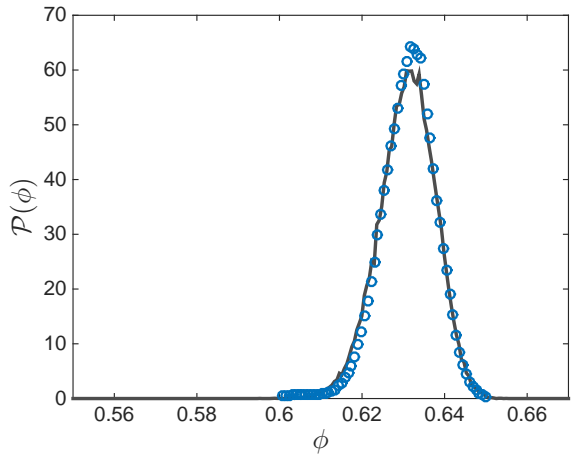}
\caption{Distribution of jammed packing fractions
$\mathcal{P}_1(\phi)$ from protocol $1$ for $N=64$ bidisperse
spheres (solid line), compared to predictions obtained from
${\mathcal P}_1(\phi)= -M_0^{-1}dM_1(\phi)/d\phi$ with $M_1(\phi)$
given by Eq.~\ref{n0} and ${\mathcal F}(\phi)S_1(\phi) l_1(\phi)$
given by Eq.~\ref{average_strain2} using the measured value of
${\overline \gamma}_j$ for $N = 64$ bidisperse spheres (circles).}
\label{fig:supp_one}
\end{figure}

Here, we confirm that we can calculate the distribution ${\mathcal
  P}_1(\phi)$ of jammed packing fractions (for isotropic compression)
using data from protocol $2$. In Fig.~\ref{fig:supp_one}, we show that
${\mathcal P}_1(\phi)$ from protocol $1$ agrees with ${\mathcal P}_1(\phi)=
-M_0^{-1}dM_1(\phi)/d\phi$ with
\begin{equation}
\label{an0}
M_1(\phi)=M_0 \exp\left[-\int^\phi_{\phi_0} {\mathcal F}(\phi')S_1(\phi') \ell_1(\phi')d\phi'\right],
\end{equation}  
and the product ${\mathcal F}S_1 \ell_1$ given by the 
measured value of ${\overline \gamma}_j$,
\begin{equation}
\label{a_average_strain2}
{\overline \gamma}_j  (\phi) \simeq \frac{\alpha}{ {\mathcal F}(\phi) S_1(\phi) \ell^\beta_1(\phi)}.
\end{equation} 
We find that $\beta \approx 1.75$ in 3D, whereas $\beta \approx 2.0$
in 2D.

\section{Stress anisotropy in frictionless packings}
\label{stress}

In this Appendix, we show the typical structure of the geometrical families 
for frictionless packings and measure the ratio of the stress anisotropy 
to the pressure for these packings. 
When deforming granular packings, the deformation method can be either
strain- or stress-controlled. In strain-controlled deformations, a
strain is applied to the system and the resulting stress is measured.
In contrast, in stress-controlled deformations, a stress is applied to
the system, and the resulting strain is measured. In simulations with
periodic boundary conditions, one of the simplest deformation methods
is the application of simple shear strain $\gamma$ using Lees-Edwards
boundary conditions.  Thus, Lees-Edwards simple shear is
strain-controlled. During the applied simple shear strain, one can
measure the resulting stress of the system.

\begin{figure}[!h]
\includegraphics[width=8.5cm]{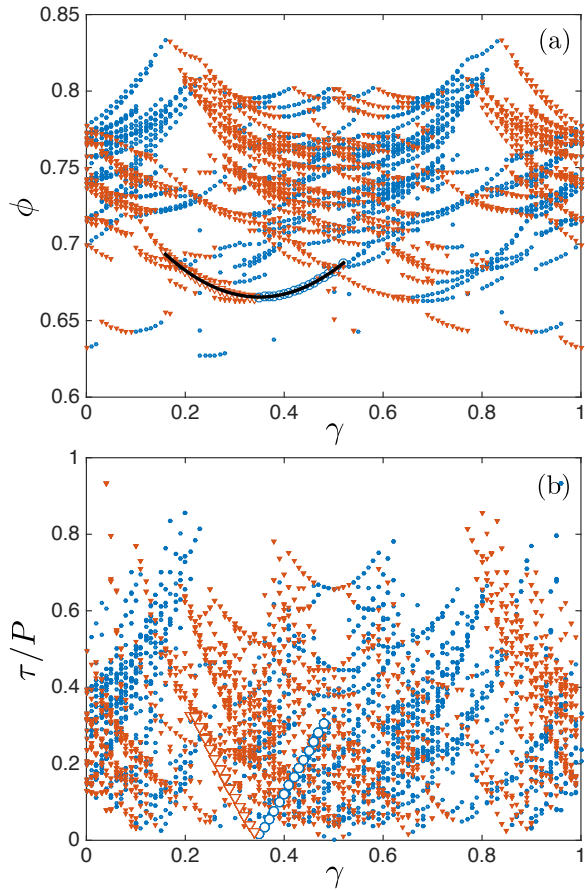}
\caption{(a) Packing fraction $\phi$ and (b) ratio of the stress anisotropy $\tau$ to the pressure $P$ versus shear strain $\gamma$ for
all isostatic jammed $N=6$ bidisperse disk packings. The solid black line 
in (a) obeys $\phi =
A(\gamma-\gamma_0)^2 + \phi_0$ with $A=0.776$, $\phi_0=0.665$, and
$\gamma_0=0.35$. Filled circles (downward triangles) indicate
packings with positive (negative) local slope of $\phi$ versus $\gamma$.}
\label{fig:supp_two}
\end{figure}

To measure the stress in 2D, we define the $2\times 2$ stress tensor 
\begin{equation}
\Sigma_{\lambda \delta} = \frac{1}{L^2} \sum_{i>j} f_{ij\lambda} r_{ij\delta},
\end{equation}
where $f_{ij\lambda}$ is the $\lambda$-component of the pairwise
repulsive force ${\vec f}_{ij}$ on particle $i$ from particle $j$,
$r_{ij\delta}$ is the $\delta$-component of the center-to-center
distance vector ${\vec r}_{ij}$ between particles $i$ and $j$, $\lambda=x,y$, and
$\delta=x,y$. In Fig.~\ref{fig:supp_two} (b), we show the ratio of the
stress anisotropy $\tau = |\Sigma_1 - \Sigma_2|/2$ to the pressure
$P=(\Sigma_1+\Sigma_2)/2$, where $\Sigma_1$ and $\Sigma_2$ are the two
eigenvalues of the stress tensor, as a function of the shear strain
$\gamma$.  In many cases the normalized stress anisotropy $\tau/P$
follows nearly linear segments along the geometrical families, which
appear as parabolas when $\phi$ for each jammed packing is plotted
versus $\gamma$ (Fig.~\ref{fig:supp_two} (a)).  However, for other
geometrical families, $\tau/P$ appears quadratic in $\gamma$.  For both
cases, $\tau/P$ decreases when $-|d\phi/d\gamma| \le 0$ and increases
when $d\phi/d\gamma \ge 0$ along each geometrical family.

\section{Comparison of pure and simple shear}
\label{pure_shear}

In this Appendix, we compare results for simple shear and pure shear protocols. 
All of the results for protocol $2$ presented in the main text were
obtained using simple shear strain using Lees-Edwards boundary
conditions. We have also studied strain-induced jamming using pure
shear, where the separation between one pair of opposing edges of the
simulation box is increased by $1+\gamma$ and the separation between
the other pair of opposing edges is decreased by $1/(1+\gamma)$.  This
deformation is the simplest example of a variable-shape simulation
cell method that conserves volume.

\begin{figure}[!h]
\includegraphics[width=8.5cm]{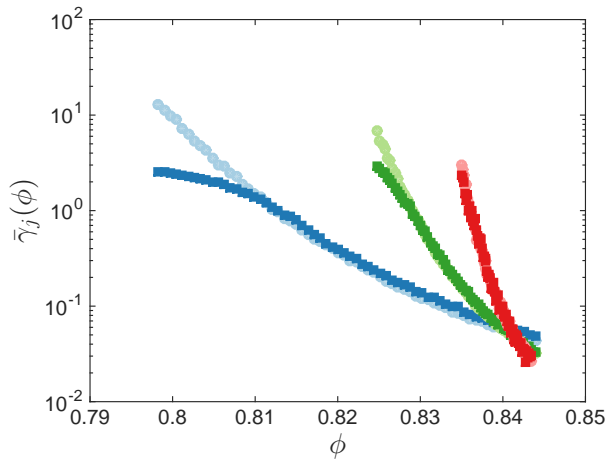}
\caption{Average shear strain ${\overline \gamma}_j(\phi)$
required to jam an initially unjammed configuration at $\phi$ 
using simple shear (light circles) and pure shear strain 
(dark squares) for $N = 32$ (blue symbols), $128$ (green symbols),
and $512$ (red symbols). }
\label{fig:supp_three}
\end{figure}

We show in Fig.~\ref{fig:supp_three} that the strain ${\overline
  \gamma}_j(\phi)$ required to jam an initially unjammed configuration
at packing fraction $\phi$ behaves qualitatively the same for packings
generated via simple and pure shear strain. We see that the results for
${\overline \gamma}_j$ for simple and pure shear strain begin to deviate at low
$\phi$, but the deviation decreases with increasing system size.  In the 
studies of pure shear, we stopped the simulations when the size 
of the cell in the thin direction caused interactions between
a disk in the main cell and one of its own periodic images.  If the system 
had not yet jammed, we did not include this trial in the measurement 
of ${\overline \gamma}_j$.  Thus, in the pure shear simulations, 
our results at low $\phi$ and small system sizes were biased toward 
small strains.  This effect vanishes in the large-system limit.      

\section{Theoretical model}
\label{theory}

In this Appendix, we elaborate some of the key aspects of the
theoretical model described in the main text. We develop the
theoretical model using an analogy with absorption problems to
calculate the probability to obtain isostatic jammed packings as a
function of the path that the system traverses in configuration
space. In principle, the number density of jammed packings ${\mathcal
  F}$, depends on the $2N$ coordinates of the disks, the packing
fraction $\phi$, and boundary deformation $\gamma$, but not the
packing-generation protocol.  After integrating over the $2N$
coordinates of the disks, ${\mathcal F}$ is a function of $\phi$ and
$\gamma$.  However, for the theoretical description, we assume that
the number density ${\mathcal F}(\phi)$ is only a function of packing
fraction since ${\mathcal F}$ becomes independent of $\gamma$ in the
large system limit as shown in Fig.~2 (c) in the main text.

We imagine that a one-dimensional trajectory
$\mathcal{L}(\phi,\gamma)$ through configuration space will encounter
the basin of attraction of a jammed packing with a probability
${\mathcal F}(\phi)S(\phi) d\mathcal{L}$ during a step of size $d
\mathcal{L}$ in configuration space, where $S(\phi)$ is the
$2N-1$-dimensional cross-section of the basin of attraction of a
jammed packing perpendicular to $d \mathcal{L}$ and $d\mathcal{L} = 
\ell_1(\phi) d\phi$ and $\ell_2(\phi) d\gamma$ for protocols $1$ and $2$, 
respectively.  

For the results presented in this manuscript, our calculations do not
require complete enumeration of jammed packings and independent
measurements of ${\mathcal F}(\phi)$, $S(\phi)$, and $\ell(\phi)$.  An
advantage of our work is that we showed that one can obtain the
product ${\mathcal F}S\ell$ without complete enumeration by measuring
the decrease in the number of unjammed configurations during shear.
We showed (for fast quenching protocols) that the product ${\mathcal
  F}S\ell$ depends only on the packing fraction $\phi$, and not on the shear
strain $\gamma$. This result implies that we can use the shear protocol to
predict the distribution of jammed packings obtained from the
isotropic compression protocol.  

To obtain $\ell_{1,2}$, we measured the cumulative distance traveled
by the system in configuration space after taking a step in packing
fraction $d \phi$ (Protocol 1) or a step in shear strain $d \gamma$
(Protocol 2) and minimizing the total potential energy: 
\begin{equation}
\ell_{1,2}(\phi) = \sqrt{\sum_{i=1}^N |\delta \vec{r}_i|^2},
\end{equation}
where $\delta \vec{r}_i$ is the change in position of particle $i$
following the compression or shear step and subsequent energy
minimization.  We averaged $\ell_{1,2}(\phi)$ over at least 
$100$ independent trajectories.

\end{document}